\documentclass[10pt,preprintnumbers,showpacs]{revtex4}
\usepackage{graphicx, epsfig}

\def\xslash#1{{\rlap{$#1$}/}}

\def\beq{\begin{equation}}
\def\eeq{\end{equation}}
\def\beqa{\begin{eqnarray}}
\def\eeqa{\end{eqnarray}}
\def\iar{\begin{array}{l}}
\def\ear{\end{array}}

\begin{document}

%\preprint{BIHEP-TH-2003-33}
\title{Gauge dependence of fermion mass renormalization schemes}
\author{Yong Zhou}
\affiliation{Institute of High Energy Physics, Academia Sinica, P.O. Box 918(4), Beijing 100049, China, Email: zhouy@ihep.ac.cn}

\begin{abstract}
We discuss the gauge dependence of fermion mass definition and physical result under the conventional on-shell mass renormalization scheme and the recently proposed pole mass renormalization scheme in standard model. By the two-loop calculations of top quark's mass counterterm and the cross section of physical process $W^{-}W^{+}\rightarrow b\bar{b}$ we strictly prove that the on-shell mass renormalization scheme makes fermion mass definition and the physical result gauge dependent at the first time, while we don't find such gauge dependence in the pole mass renormalization scheme. Besides, our calculation also shows the difference of the physical result between the two mass renormalization schemes cannot be neglected at two-loop level. 
\end{abstract}

%11.10.Gh: Renormalization
%12.15.Ff: Quark and lepton masses and mixing
%12.15.Lk: Electroweak radiative corrections
\pacs{11.10.Gh, 12.15.Ff, 12.15.Lk}
\maketitle

\section{Introduction}

Along with the improvement of the measurement precision of physical experiments, we need more and more radiative-correction calculations beyond one-loop level. The fermion mass renormalization scheme has been discussed enough at one-loop level \cite{c1,cin0}, but is rarely discussed beyond one-loop level. At present there are two mass renormalization schemes: the conventional on-shell scheme \cite{c1,cin0} and the recently proposed pole scheme \cite{c2}. People have pointed out that the mass definition of pole scheme is gauge invariant compared with the on-shell scheme \cite{c2,c3}, but this conclusion is only based on a native hypothesis that the location of the complex pole of the propagator is gauge independent at any order in perturbation theory \cite{cin} and a strict proof of the hypothesis hasn't been present. So we need to further investigate this problem. In this paper we will investigate the gauge dependence of fermion mass definition.

In order to obtain the fermion mass renormalization scheme beyond one-loop level we firstly investigate the fermion inverse propagator
\beq
  i S^{-1}(\xslash p)\,=\,\xslash p-m_0+\Sigma(\xslash p)\,\equiv\,
  \xslash p(a\gamma_L+b\gamma_R)+c\gamma_L+d\gamma_R\,,
\eeq
where $m_0$ is the bare fermion mass, $\gamma_L$ and $\gamma_R$ are the left- and right- handed helicity operators, and
\beq
  \Sigma(\xslash p)\,=\,\xslash p\gamma_L\,\Sigma^L(p^2)+\xslash p\gamma_R\,\Sigma^R(p^2)
  +m\,\gamma_L\Sigma^{S,L}(p^2)+m\,\gamma_R\Sigma^{S,R}(p^2)\,.
\eeq
When we renormalize fermion mass, the fermion propagator should be changed into \cite{c1,c4}
\beq
  S(\xslash p)\,=\,\frac{\xslash p(a\gamma_L+b\gamma_R)-d\gamma_L-c\gamma_R}
  {p^2 a b - c d}\,=\,\frac{i(m_0+\xslash p\gamma_L(1+\Sigma^L)
  +\xslash p\gamma_R(1+\Sigma^R)-m\gamma_L\Sigma^{S,R}-m\gamma_R\Sigma^{S,L})}
  {p^2(1+\Sigma^L)(1+\Sigma^R)-(m_0-m\Sigma^{S,L})(m_0-m\Sigma^{S,R})}\,.
\eeq
The on-shell mass renormalization scheme requires the real part of the propagator's denominator equal to zero at the physical mass point, i.e. \cite{c4}
\beq
  Re\,[m^2(1+\Sigma^L(m^2))(1+\Sigma^R(m^2))-(m+\delta m-m\,\Sigma^{S,L}(m^2))
  (m+\delta m-m\,\Sigma^{S,R}(m^2))]\,=\,0\,.
\eeq
Solving this equation about $\delta m$ to two-loop level we obtain
\beqa
  \delta m&=&\frac{m}{2}Re[\Sigma^L(m^2)+\Sigma^R(m^2)+\Sigma^{S,L}(m^2)
  +\Sigma^{S,R}(m^2)]-\frac{m}{8}Re^2 [\Sigma^L(m^2)-\Sigma^R(m^2)] \nonumber \\
  &+&\frac{m}{2}\bigl{(} Im\Sigma^{S,L}(m^2)Im\Sigma^{S,R}(m^2)
  -Im\Sigma^L(m^2)Im\Sigma^R(m^2) \bigr{)}\,.
\eeqa
In Eq.(5) we have used the fact $\Sigma^{S,L}=\Sigma^{S,R}$ at one-loop level.

Different from the on-shell scheme, the pole mass renormalization scheme requires both the real part and the imaginary part of the propagator's denominator equal to zero at the location of the propagator's complex pole \cite{cin2}
\beq
  \bar{s}(1+\Sigma^L(\bar{s}))(1+\Sigma^R(\bar{s}))
  -(m_2+\delta m_2-m_2\Sigma^{S,L}(\bar{s}))(m_2+\delta m_2-m_2\Sigma^{S,R}(\bar{s}))
  \,=\,0\,,
\eeq
where $m_2$ is the physical mass and $\bar{s}=m_2^2-i m_2\Gamma_2$ is the location of the propagator's complex pole. Solving this equation about $\delta m_2$ to two-loop level we obtain
\beqa
  \delta m_2&=&\frac{m_2}{2}Re[\Sigma^L(m_2^2)+\Sigma^R(m_2^2)+\Sigma^{S,L}(m_2^2)
  +\Sigma^{S,R}(m_2^2)]-\frac{m_2}{8}Re^2 [\Sigma^L(m_2^2)-\Sigma^R(m_2^2)] \nonumber \\
  &+&\frac{m_2}{2}\bigl{(} Im\Sigma^{S,L}(m_2^2)Im\Sigma^{S,R}(m_2^2)
  -Im\Sigma^L(m_2^2)Im\Sigma^R(m_2^2) \bigr{)} \nonumber \\
  &+&\frac{\Gamma_2}{2}\bigl{(} m_2^2\,Im[\Sigma^{L\prime}(m_2^2)+\Sigma^{R\prime}(m_2^2)
  +\Sigma^{S,L\prime}(m_2^2)+\Sigma^{S,R\prime}(m_2^2)]
  +Im[\Sigma^L(m_2^2)+\Sigma^R(m_2^2)] \bigr{)}\,,
\eeqa
where $\Sigma^{L\prime}(m_2^2)=\partial\Sigma^L(m_2^2)/\partial p^2$ et al and the one-loop fermion decay width $\Gamma_2$ is
\beq
  \Gamma_2\,=\,m_2\,Im[\Sigma^L(m_2^2)+\Sigma^R(m_2^2)+\Sigma^{S,L}(m_2^2)
  +\Sigma^{S,R}(m_2^2)]\,.
\eeq

In order to investigate the gauge dependence of the two fermion mass definitions we calculate the gauge dependence of the branch cut of top quark's two-loop mass counterterm in the following section; in section III we calculate the cross section of the physical process $W^{-}W^{+}\rightarrow b\bar{b}$, i.e. a pair of gauge boson W decaying into a pair of bottom quarks, to investigate the gauge dependence of physical result under the on-shell mass renormalization scheme; lastly we give the conclusion in section IV.

\section{Gauge dependence of fermion mass definition in the on-shell and pole mass renormalization schemes}

If fermion mass is defined gauge independently, its counterterm must be also gauge independent \cite{c5}. This can be used to test whether the fermion mass definition is gauge independent. In the following calculations we will investigate whether Eq.(5) and Eq.(7) are gauge independent. For convenience we only consider W gauge parameter's dependence in the $R_{\xi}$ gauge and only introduce physical parameter's counterterms. Note that we have used the computer program packages {\em FeynArts} and {\em FeynCalc} \cite{c6} in the following calculations.

We give top quark's mass counterterm as an example to investigate this problem. Since nobody has discussed this problem before, we begin the discussion from the one-loop level. The one-loop top quark's $\xi_W$-dependent self-energy and tadpole diagrams are shown in Fig.1 and Fig.2, where $\xi_W$ is W gauge parameter. We note that in order to investigate the gauge dependence of mass counterterm the tadpole diagrams of Fig.2 must be included \cite{cin}.
\begin{figure}[htbp]
\begin{center}
  \epsfig{file=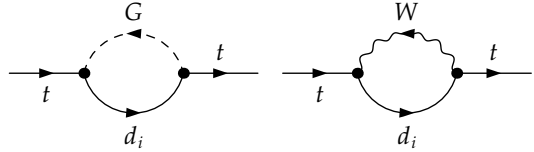, width=7cm} \\
  \caption{Diagrams of one-loop top quark's $\xi_W$-dependent self energy.}
\end{center}
\end{figure}
\begin{figure}[htbp]
\begin{center}
  \epsfig{file=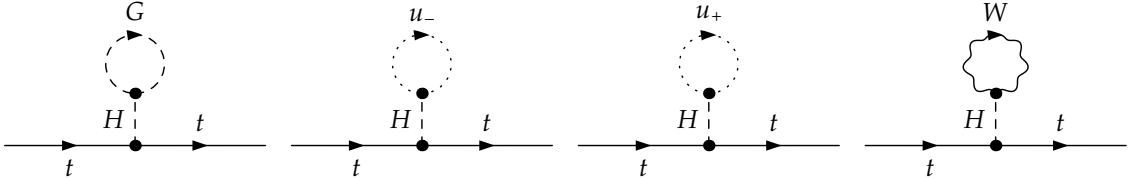, width=15cm} \\
  \caption{Tadpole diagrams of one-loop top quark's $\xi_W$-dependent self energy.}
\end{center}
\end{figure}
We firstly obtain the $\xi_W$-dependent contribution of Fig.1 to top quark's one-loop mass counterterm (see Eqs.(5,7))
\beq
  \frac{m_t}{2}Re[\Sigma^L(m_t^2)+\Sigma^R(m_t^2)+\Sigma^{S,L}(m_t^2)
  +\Sigma^{S,R}(m_t^2)]_{\xi_W}\,=\,\frac{e^2 m_t}{64\pi^2 s_w^2}\xi_W
  \bigl{(} 1+B_0(0, \xi_W m_W^2, \xi_W m_W^2) \bigr{)}\,,
\eeq
where the subscript $\xi_W$ denotes the $\xi_W$-dependent part of the quantity, $m_t$ and $m_W$ are the masses of top quark and gauge boson W, $e$ is electron charge, $s_w$ is the sine of weak mixing angle, and $B_0$ is the Passarino-Veltman function \cite{c7}. On the other hand, the $\xi_W$-dependent contribution of Fig.2 to top quark's one-loop mass counterterm is
\beq
  \frac{m_t}{2}Re[\Sigma^L(m_t^2)+\Sigma^R(m_t^2)+\Sigma^{S,L}(m_t^2)
  +\Sigma^{S,R}(m_t^2)]_{\xi_W}\,=\,-\frac{e^2 m_t}{64\pi^2 s_w^2}\xi_W
  \bigl{(} 1+B_0(0, \xi_W m_W^2, \xi_W m_W^2) \bigr{)}\,.
\eeq
From Eqs.(9,10)  one readily see that the top quark's one-loop mass counterterm is gauge independent.

Now we calculate the gauge dependence of top quark's two-loop mass counterterm. For convenience we only calculate the gauge dependence of the branch cut of top quark's mass counterterm. Since the gauge-dependent branch cut of top quark's mass counterterm comprises gauge-dependent Heaviside function only (see below) and the other part of top quark's mass counterterm doesn't contain such function, this simplified treatment is reasonable and will not break our final conclusion. The branch cut of top quark's two-loop mass counterterm will be calculated by the {\em cutting rules} \cite{c8,c9}. The concrete algorithm used here can be found in Ref.\cite{c9}. From Eqs.(5,7) we find at two-loop level the top quark's mass counterterms are different in the two mass renormalization schemes. We firstly discuss the case of on-shell mass renormalization scheme. It is obvious that at two-loop level the second term of the r.h.s. of Eq.(5) doesn't contain branch cut. So we don't need to consider this term. Through the imaginary
part of Fig.1 we obtain the $\xi_W$-dependent branch cut of the last term of the r.h.s. of Eq.(5) \beqa
  &&\frac{m_t}{2}\bigl{[} Im\Sigma^{S,L}(m_t^2)Im\Sigma^{S,R}(m_t^2)
  -Im\,\Sigma^L(m_t^2)Im\,\Sigma^R(m_t^2) \bigr{]}_{\xi_W-cut} \nonumber \\
  &=&-\frac{e^4\,m_W}{8192\pi^2\,s_w^4\,x_t^{5/2}}\sum_{i=d,s,b}|V_{3i}|^2
  A_i\,C_i\sum_{j=d,s,b}|V_{3j}|^2 B_j(x_t-\xi_W+ x_j)\,\theta[m_t-m_j-\sqrt{\xi_W}m_W]
  \nonumber \\
  &+&\frac{e^4\,m_W}{8192\pi^2\,s_w^4\,x_t^{3/2}}\sum_{i,j=d,s,b}|V_{3i}|^2
  |V_{3j}|^2  B_i\,B_j \bigl{(} (x_t-\xi_W)(x_t-\xi_W-3 x_i)+(x_t-\xi_W+x_i)x_j \bigr{)}
  \nonumber \\
  &\times&\theta[m_t-m_i-\sqrt{\xi_W}m_W]\,\theta[m_t-m_j-\sqrt{\xi_W}m_W]\,,
\eeqa
where the subscript $\xi_W\hspace{-1mm}-\hspace{-1mm}cut$ denotes the $\xi_W$-dependent branch cut of the quantity, $V_{3i}$ and $V_{3j}$ are CKM matrix elements \cite{c10}, $m_i$ and $m_j$ are the masses of down-type $i$ and $j$ quarks, $x_t=m_t^2/m_W^2$, $x_i=m_i^2/m_W^2$ and $x_j=m_j^2/m_W^2$, $\theta$ is the Heaviside function, and
\beqa
  A_i&=&(x_t^2-2(x_i+1)x_t+(x_i-1)^2)^{1/2}\,, \nonumber \\
  B_i&=&(x_t^2-2(\xi_W+x_i)x_t+(\xi_W-x_i)^2)^{1/2}\,, \nonumber \\
  C_i&=&x_t^2+(1-2 x_i)x_t+x_i^2+x_i-2\,.
\eeqa

In order to investigate the gauge dependence of the branch cut of the first term of the r.h.s. of Eq.(5) we need to calculate top quark's two-loop self energy. The top quark's two-loop self energy can be classified into two kinds: one contains the one-loop counterterm, the second doesn't contain any counterterm. Since except for CKM matrix elements all the one-loop physical parameter's counterterms are real numbers and free of branch cut \cite{cin0}, the first kind self energy doesn't contribute to the branch cut of Eq.(5), because except the one-loop counterterm the left part of the self energy is the one-loop self energy which real part is free of branch cut. Here we don't need to worry about the problem of the CKM matrix elements and their counterterms being complex numbers, because the total contribution of them to Eq.(5) is real number (see the appendix for example). So we only need to consider the second kind self energy. According to the {\em cutting rules} of Ref.\cite{c9}, the second kind self energy can also be classified into three kinds: one doesn't contain branch cut, the second contains branch cut, but its branch cut doesn't contribute to Eq.(5), the third contains branch cut and its branch cut contributes to Eq.(5) \cite{c11}. Obviously only the third kind self energy needs to be calculated. The topologies of it are shown in Fig.3 \cite{c11}.
\begin{figure}[htbp]
\begin{center}
  \epsfig{file=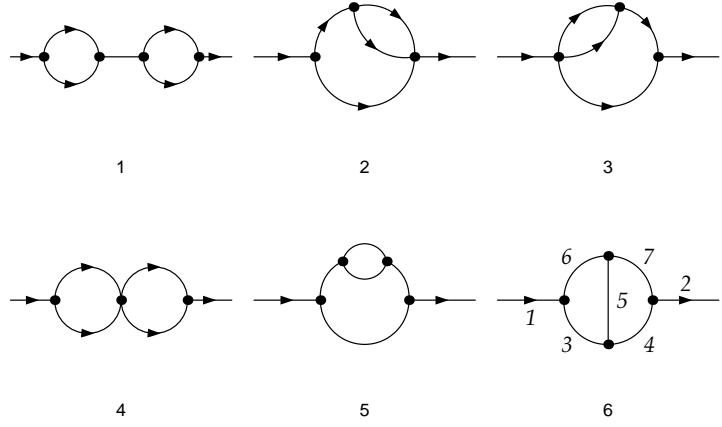, width=9.5cm} \\
  \caption{Topologies of two-loop self energy without counterterm
  which branch cuts contribute to Eq.(5).}
\end{center}
\end{figure}
One notices in Fig.3 that an one-particle-reducible diagram has been included. Although a self energy is usually defined to comprise one-particle-irreducible diagrams only, no physical principle demonstrates the one-particle-reducible diagram can not be included in a correct self energy. Furthermore in the following calculations of top quark's mass counterterm one will see the one-particle-reducible diagram in Fig.3 guarantees the gauge independence of the branch cut of top quark's mass definition under the ple mass renormalization scheme. In Fig.3 we also draw the possible cuts of the first four topologies which contribute to Eq.(5), where the propagator with an arrow on it denotes it is cut and the arrow represents the {\em positive-on-shell} momentum propagating direction of the cut propagator \cite{c9}). The possible cuts of the left two topologies which contribute to Eq.(5) are shown in Fig.4 and Fig.5.
\begin{figure}[htbp]
\begin{center}
  \epsfig{file=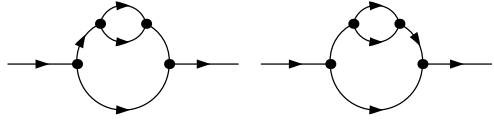, width=6.5cm} \\
  \caption{Possible cuts of the 5th topology of Fig.3
  which contribute to Eq.(5).}
\end{center}
\end{figure}
\begin{figure}[htbp]
\begin{center}
  \epsfig{file=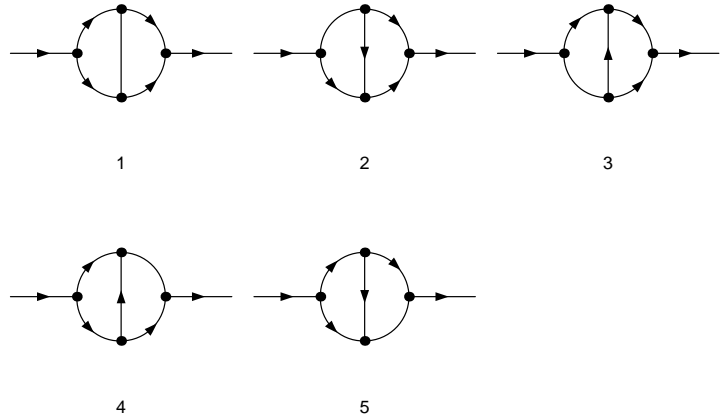, width=9.5cm} \\
  \caption{Possible cuts of the 6th topology of Fig.3
  which contribute to Eq.(5).}
\end{center}
\end{figure}
The concrete calculations of the $\xi_W$-dependent contribution of the possible cuts of Fig.3 to Eq.(5) are shown in the appendix. Here we list the final result
\beqa
  &&\frac{m_t}{2}Re[\Sigma^L(m_t^2)+\Sigma^R(m_t^2)+\Sigma^{S,L}(m_t^2)
  +\Sigma^{S,R}(m_t^2)]_{\xi_W-cut} \nonumber \\
  &=&\frac{e^4 m_W}{4096\pi^2\,s_w^4\,x_t^{5/2}}\sum_{i=d,s,b}|V_{3i}|^2
  A_i\,C_i\sum_{j=d,s,b}|V_{3j}|^2 B_j(x_t-\xi_W- x_j)\,\theta[m_t-m_j-\sqrt{\xi_W}m_W]
  \nonumber \\
  &-&\frac{e^4\,m_W}{8192\pi^2\,s_w^4\,x_t^{3/2}}\sum_{i,j=d,s,b}|V_{3i}|^2
  |V_{3j}|^2  B_i\,B_j (x_t-\xi_W-x_i)(x_t-\xi_W-x_j) \nonumber \\
  &\times&\theta[m_t-m_i-\sqrt{\xi_W}m_W]\,\theta[m_t-m_j-\sqrt{\xi_W}m_W]\,.
\eeqa

From Eq.(5) and Eqs.(11,13) we obtain the gauge dependence of the branch cut of top quark's two-loop mass counterterm under the on-shell mass renormalization scheme 
\beq
  \delta m_t|_{\xi_W-cut}\,=\,\frac{e^4\,m_W}{8192\pi^2\,s_w^4\,x_t^{5/2}}
  \sum_{i=d,s,b}|V_{3i}|^2 A_i\,C_i\sum_{j=d,s,b}|V_{3j}|^2 B_j(x_t-\xi_W-3 x_j)\,
  \theta[m_t-m_j-\sqrt{\xi_W}m_W]\,.
\eeq
According to the above discussion above Eqs.(11) Eq.(14) manifests top quark's two-loop mass counterterm is gauge dependent under the on-shell mass renormalization prescription. This means the conventional on-shell mass definition is gauge dependent according to the discussion at the begining of this section.

Then we calculate the gauge dependence of top quark's two-loop mass counterterm under the pole mass renormalization scheme. The result of the first three terms of Eq.(7) is same as the result of Eq.(5). So we only need to calculate the last term of Eq.(7). The concrete result is 
\beqa
  &&\frac{\Gamma_t}{2}\bigl{(} m_t^2\,Im[\Sigma^{L\prime}(m_t^2)+\Sigma^{R\prime}(m_t^2)
  +\Sigma^{S,L\prime}(m_t^2)+\Sigma^{S,R\prime}(m_t^2)]
  +Im[\Sigma^L(m_t^2)+\Sigma^R(m_t^2)] \bigr{)}|_{\xi_W-cut} \nonumber \\
  &=&-\frac{e^4\,m_W}{8192\pi^2\,s_w^4\,x_t^{5/2}}\sum_{i=d,s,b}|V_{3i}|^2
  A_i\,C_i\sum_{j=d,s,b}|V_{3j}|^2 B_j(x_t-\xi_W-3 x_j)\,
  \theta[m_t-m_j-\sqrt{\xi_W}m_W]\,.
\eeqa
From Eq.(7) and Eqs.(11,13,15) we readily have
\beq
  \delta m_t|_{\xi_W-cut}\,=\,0\,.
\eeq
This means in the pole mass definition we don't find the gauge dependence present in the on-shell mass definition.

\section{Gauge dependence of physical result under the on-shell mass renormalization scheme}

From the above discussion one readily sees that the conventional on-shell mass renormalization scheme is unreasonable. In fact the on-shell mass renormalization condition of Eq.(4) doesn't exist, because it requires the bare fermion mass $m_0$ is gauge dependent which is disallowed. In this section we will further discuss the effect of the gauge dependence of the on-shell fermion mass definition on physical result. We give the physical process $W^{-}W^{+}\rightarrow b\bar{b}$ as an example to illustrate this problem. Note that we only consider the gauge dependence of the branch cut containing the function $\theta[m_t-m_{down-quark}-\sqrt{\xi_W}m_W]$ of the two-loop cross section of the physical process. For convenience we hypothesize the Higgs mass and the total energy of the incoming particles are less than top quark's mass. This will not break the following conclusion. Under the hypothesis only the diagram containing a two-loop $\delta m_t$ needs to be consider.
\begin{figure}[htbp]
\begin{center}
  \epsfig{file=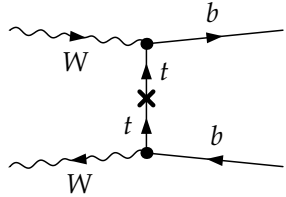, width=3.7cm} \\
  \caption{Diagram of two-loop $W^{-}W^{+}\rightarrow b\bar{b}$ containing a two-loop
  $\delta m_t$.}
\end{center}
\end{figure}
The contribution of Fig.6 to the two-loop amplitude $W^{-}W^{+}\rightarrow b\bar{b}$ is
\beq
  -\frac{e^2|V_{33}|^2 m_t\delta m_t}{s_w^2(m_b^2+m_W^2-2 p_2\cdot q_2-m_t^2)^2}
  \bigl{[} \bar{u}(q_1)\cdot{\xslash \epsilon_1}\cdot{\xslash \epsilon_2}\cdot
  {\xslash p_2}\cdot\gamma_L\cdot\nu(q_2)+2\bar{u}(q_1)\cdot{\xslash \epsilon_1}
  \cdot\gamma_L\cdot\nu(q_2)\epsilon_2\cdot q_2+m_b\bar{u}(q_1)\cdot{\xslash \epsilon_1}
  \cdot{\xslash \epsilon_2}\cdot\gamma_R\cdot\nu(q_2) \bigr{]}\,,
\eeq
where $m_b$ is bottom quark's mass, $p_2$, $q_1$ and $q_2$ are the momentums of $W^{+}$, bottom and anti-bottom quarks, and $\epsilon_1$ and $\epsilon_2$ are the polarization vectors of $W^{-}$ and $W^{+}$. The contribution of Eq.(17) to two-loop $|\cal{M}$$(W^{-}W^{+}\rightarrow b\bar{b}|^2$ is
\beqa
  &&\frac{e^4 |V_{33}|^2 m_t\delta m_t}{27m_W^2 s_w^4(x_t+2 x_p-1)^2}\Bigl{[}
  -3\bigl{(} x_E(2 x_p-1)(2 x_p+1)^2-4 x_p^2(4(x_p-1)x_p+5) \bigr{)}\sum_{i=u,c,t}
  \frac{|V_{i3}|^2}{2 x_p+x_i-1} \nonumber \\
  &+&\frac{c_w^2(3 x_E+2)-2}{x_E(c_w^2 x_E-1)}\bigl{(} -8(x_E-2)x_p^3
  +4(x_E-3)(x_E+2)x_p^2+4 x_E^2 x_p-3 x_E(x_E+2) \bigr{)} \Bigr{]}\,,
\eeqa
where $x_p=p_2\cdot q_2/m_W^2$ and $x_E=E_{cm}^2/m_W^2$ ($E_{cm}$ is the total energy of the incoming particles in the center of mass frame). For convenience in Eq.(18) we have approximated $m_b^2/m_W^2\rightarrow 0$ and averaged the result over the incoming particles' polarization vectors and summed the result over the outgoing particles' helicity states.

Since there is no other origin of function $\theta[m_t-m_{down-quark}-\sqrt{\xi_W}m_W]$ in the two-loop $|\cal{M}$$(W^{-}W^{+}\rightarrow b\bar{b}|^2$, Eq.(14) and Eq.(18) mean the on-shell mass renormalization scheme makes the cross section of the physical process $W^{-}W^{+}\rightarrow b\bar{b}$ gauge dependent. From Eqs.(14,18) we also find that the gauge dependence caused by the on-shell scheme cannot be neglected at two-loop level. The concrete numerical result is shown in Fig.7.
\begin{figure}[htbp]
\begin{center}
  \epsfig{file=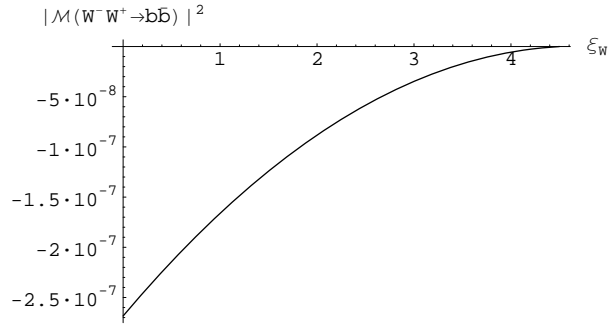, width=8cm} \\
  \caption{Gauge dependence of two-loop $|\cal{M}$$(W^{-}W^{+}\rightarrow b\bar{b})|^2$
  in on-shell mass renormalization scheme with Higgs mass 120Gev.}
\end{center}
\end{figure}
In Fig.7 we have used the data: $x_p=1$, $x_E=4$, $e=0.3028$, $s_w^2=0.2312$, $m_W=80.42Gev$, $m_u=3Mev$, $m_c=1.25Gev$, $m_t=174.3Gev$, $m_d=6Mev$, $m_s=120Mev$, $m_b=4.2Gev$, $|V_{ub}|=0.004$, $|V_{cb}|=0.040$, $|V_{td}|=0.009$, $|V_{ts}|=0.039$ and $|V_{tb}|=0.999$ \cite{c12}. Contrarily, from Eq.(16) one readily find that the branch cut containing the function $\theta[m_t-m_{down-quark}-\sqrt{\xi_W}m_W]$ of the two-loop cross section of the physical process $W^{-}W^{+}\rightarrow b\bar{b}$ is gauge independent under the pole mass renormalization scheme.

\section{Conclusion}

We have discussed the gauge dependence of fermion mass definition under the on-shell and pole mass renormalization schemes. Through the calculation of top quark's two-loop mass counterterm we strictly prove that the fermion mass definition is gauge dependent under the on-shell mass renormalization scheme at the first time. This leads to the physical result obtained by the on-shell mass renormalization scheme also gauge dependent. On the contrary, we don't find such gauge dependence in the pole mass renormalization scheme. Besides, the gauge-dependent difference of the physical result between the two mass renormalization schemes cannot be neglected at two-loop level. Therefore we should use the pole mass renormalization scheme rather than the on-shell mass renormalization scheme to calculate physical results beyond one-loop level.

\vspace{5mm} {\bf \Large Acknowledgments} \vspace{2mm}

The author thanks Prof. Xiao-Yuan Li and Prof. Cai-dian Lu for the useful guidance.

\vspace{5mm} {\bf \Large Appendix} \vspace{2mm}

In this appendix we list the $\xi_W$-dependent contributions of the possible cuts of Fig.3 to Eq.(5). We firstly calculate the $\xi_W$-dependent contribution of the possible cut of the first topology of Fig.3 to Eq.(5).
\begin{figure}[htbp]
\begin{center}
  \epsfig{file=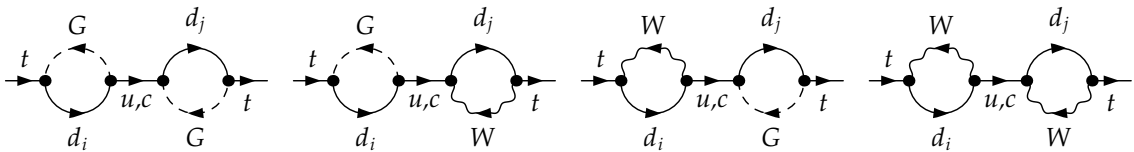, width=15cm} \\
  \caption{Diagrams of top quark's self energy of the first topology of Fig.3.}
\end{center}
\end{figure}
Using the cutting rules of Ref.\cite{c9} we obtain the $\xi_W$-dependent contribution of the cuts of Fig.8 to Eq.(5):
\beqa
  &&\frac{m_t}{2}Re[\Sigma^L(m_t^2)+\Sigma^R(m_t^2)+\Sigma^{S,L}(m_t^2)
  +\Sigma^{S,R}(m_t^2)]_{\xi_W-cut} \nonumber \\
  &=&\frac{e^4 m_W}{8192\pi^2 s_w^4 x_t^{5/2}}\sum_{i=d,s,b}|V_{3i}|^2
  B_i(x_t-\xi_W-x_i)\bigl{(} B_i x_t(x_t-\xi_W-x_i)-2 A_i\,C_i
  +2\sum_{j=d,s,b}|V_{3j}|^2 A_j\,C_j \bigr{)} \nonumber \\
  &\times&\theta[m_t-m_i-\sqrt{\xi_W}m_W]-\frac{e^4 m_W}{8192\pi^2 s_w^4 x_t^{3/2}}
  \sum_{i,j=d,s,b}|V_{3i}|^2 |V_{3j}|^2 B_i\,B_j(x_t-\xi_W-x_i)(x_t-\xi_W-x_j)
  \nonumber \\
  &\times&\theta[m_t-m_i-\sqrt{\xi_W}m_W]\,\theta[m_t-m_j-\sqrt{\xi_W}m_W]\,.
\eeqa

Since there isn't four-particle-interaction vertex including fermion in standard model, the second, 3th and 4th topologies of Fig.3 cannot contribute to Eq.(5). So we don't need to consider their contributions.

The top quark's self-energy diagrams of the 5th topology of Fig.3 which satisfy the cutting conditions of Fig.4 are shown in Fig.9.
\begin{figure}[htbp]
\begin{center}
  \epsfig{file=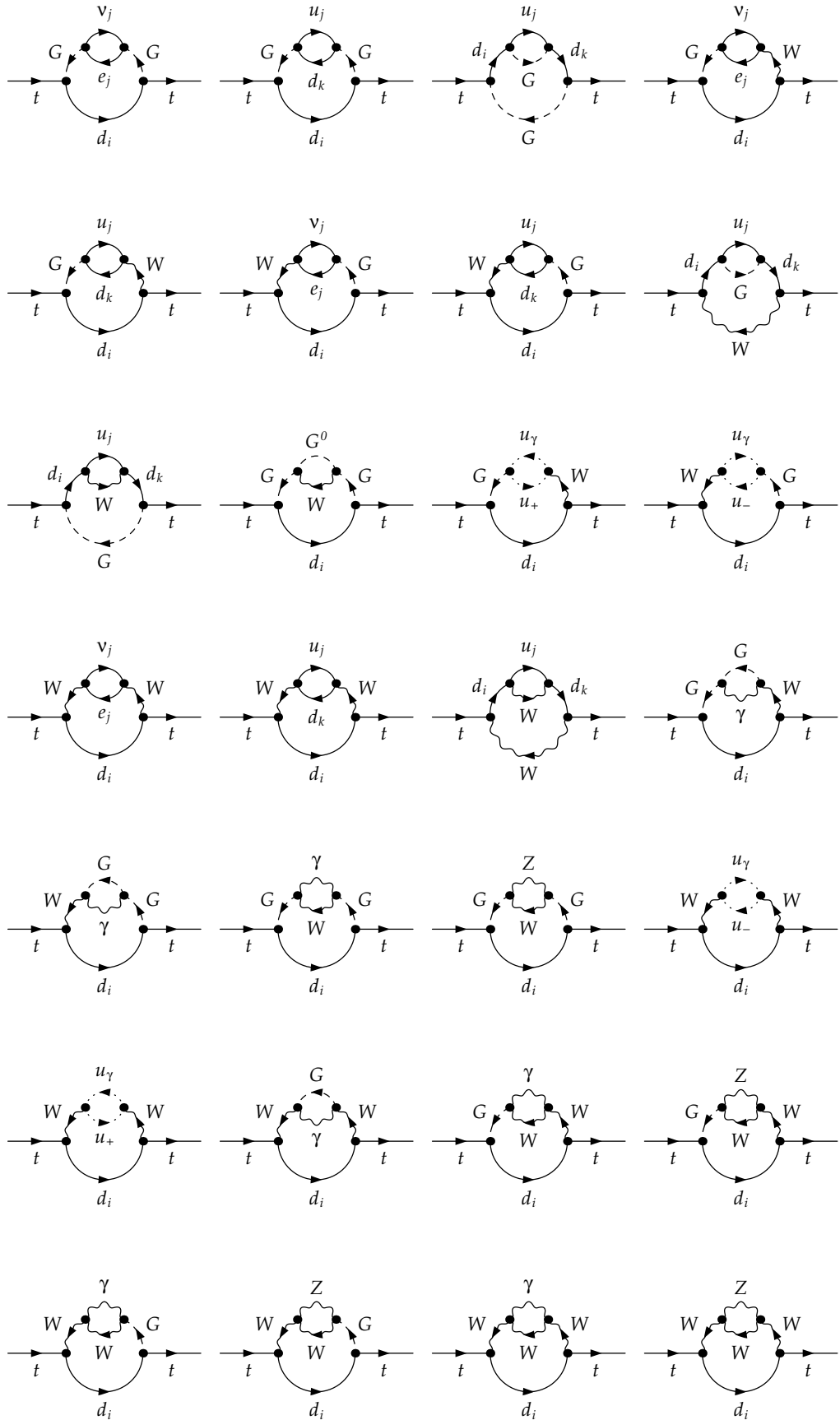, width=13.5cm} \\
  \caption{Diagrams of top quark's self energy of the 5th topology of Fig.3 which
  satisfy the cutting conditions of Fig.4.}
\end{center}
\end{figure}
Using the cutting rules we obtain the $\xi_W$-dependent contribution of the cuts of Fig.9 to Eq.(5):
\beqa
  &&\frac{m_t}{2}Re[\Sigma^L(m_t^2)+\Sigma^R(m_t^2)+\Sigma^{S,L}(m_t^2)
  +\Sigma^{S,R}(m_t^2)]_{\xi_W} \nonumber \\
  &=&-\frac{e^4 m_W}{8192\pi^2 s_w^2 x_t^{3/2}}(\xi_W^3-3\xi_W^2-9\xi_W+11)
  \theta[1-\xi_W]\sum_{i=d,s,b}|V_{3i}|^2 A_i\,C_i \nonumber \\
  &+&\frac{3 e^4 m_W}{8192\pi^2 s_w^2 x_t^{3/2}\,\xi_W^2}(\xi_W-1)^3\,
  \theta[\xi_W-1]\sum_{i=d,s,b}|V_{3i}|^2 B_i \nonumber \\
  &\times&\bigl{(} x_i^2-(2 x_t+\xi_W)x_i+x_t(x_t-\xi_W) \bigr{)}
  \theta[m_t-m_i-\sqrt{\xi_W}m_W] \nonumber \\
  &+&\frac{3 e^4 m_W}{8192\pi^2 c_w^4 s_w^4 x_t^{3/2}\,\xi_W^2}D^3\,
  \theta[\sqrt{\xi_W}-1/c_w-1]\sum_{i=d,s,b}|V_{3i}|^2 B_i \nonumber \\
  &\times&\bigl{(} x_i^2-(2 x_t+\xi_W)x_i+x_t(x_t-\xi_W) \bigr{)}
  \theta[m_t-m_i-\sqrt{\xi_W}m_W]\,,
\eeqa
where
\beq
  D\,=\,\bigl{(} (\xi_W-1)^2 c_w^4-2(\xi_W+1)c_w^2+1 \bigr{)}^{1/2}\,.
\eeq
Here we have restricted ourselves to $\xi_W>0$ \cite{c3} and used the present experiment result $m_t<m_H+m_W$ \cite{c13}.

Lastly we calculate the $\xi_W$-dependent contribution of the possible cut of the 6th topology of Fig.3 to Eq.(5).
\begin{figure}[htbp]
\begin{center}
  \epsfig{file=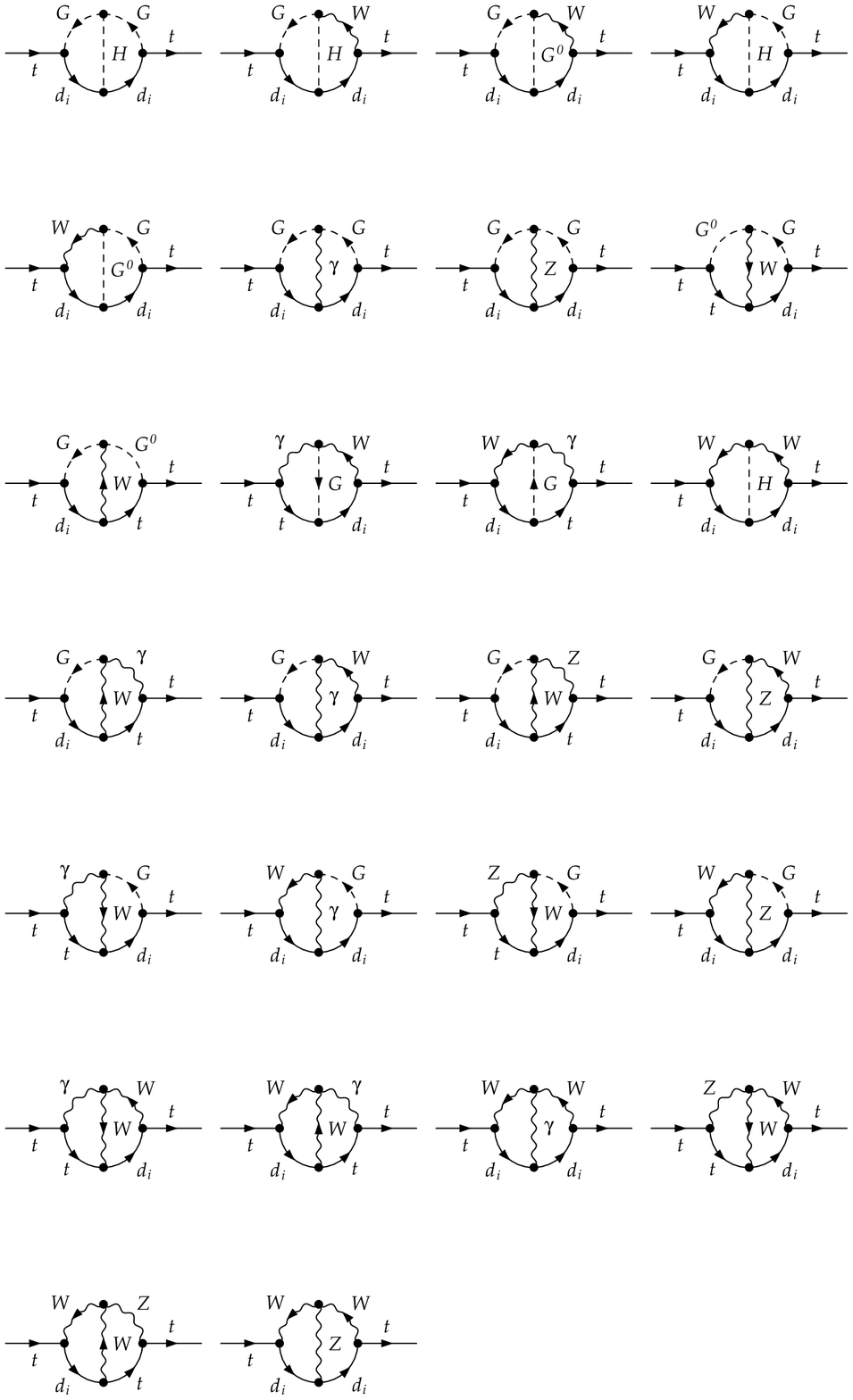, width=13.5cm} \\
  \caption{Diagrams of top quark's self energy of the 6th topology of Fig.3 which
  satisfy the cutting conditions of Fig.5.}
\end{center}
\end{figure}
We will calculate the contributions of the five cuts of Fig.5 one by one. Firstly we calculate the $\xi_W$-dependent contribution of the first cut of Fig.5 to Eq.(5). The result is
\beqa
  &&\frac{m_t}{2}Re[\Sigma^L(m_t^2)+\Sigma^R(m_t^2)+\Sigma^{S,L}(m_t^2)
  +\Sigma^{S,R}(m_t^2)]_{\xi_W-cut} \nonumber \\
  &=&\frac{e^4 m_W}{8192\pi^2 s_w^4 x_t^{5/2}}\sum_{i=d,s,b}|V_{3i}|^2
  (x_t-\xi_W-x_i)\bigl{(} 2 A_i\,B_i\,C_i-x_t(x_t-\xi_W-x_i)B_i^2 \bigr{)}
  \theta[m_t-m_i-\sqrt{\xi_W}m_W]\,.
\eeqa
We note that in the calculation of Eq.(22) we have given the photon an infinitesimal mass $\lambda$ in order to regularize the result of the diagrams the 5th propagator of which is photon (see Fig.3). The result of each such diagram contains the term $\ln\lambda$, but the total contribution of them to Eq.(5) is zero as shown in Eq.(22). Because we find that none of the diagram in Fig.10 satisfies the cutting condition of the second cut of Fig.5, the contribution of the second cut of Fig.5 to Eq.(5) is zero. Similarly, the contribution of the 4th cut of Fig.5 to Eq.(5) is also zero duo to the same reason. For the contribution of the third cut of Fig.5 to Eq.(5) we have
\beqa
  &&\frac{m_t}{2}Re[\Sigma^L(m_t^2)+\Sigma^R(m_t^2)+\Sigma^{S,L}(m_t^2)
  +\Sigma^{S,R}(m_t^2)]_{\xi_W-cut} \nonumber \\
  &=&\frac{e^4 m_W}{16384\pi^2 s_w^2 x_t^{3/2}}\sum_{i=d,s,b}|V_{3i}|^2
  A_i\,C_i(\xi_W^3-3\xi_W^2-9\xi_W+11)\theta[1-\xi_W] \nonumber \\
  &-&\frac{3 e^4 m_W}{16384\pi^2 s_w^2 x_t^{3/2}\xi_W^2}(\xi_W-1)^3\,\theta[\xi_W-1]
  \sum_{i=d,s,b}\frac{|V_{3i}|^2}{B_i}\bigl{(} x_t^4-(3\xi_W+4 x_i)x_t^3 \nonumber \\
  &+&3(\xi_W^2+x_i\xi_W+2 x_i^2)x_t^2-(\xi_W^3-2 x_i\xi_W^2-3 x_i^2\xi_W+4 x_i^3)x_t
  -(\xi_W-x_i)^3 x_i \bigr{)}\theta[m_t-m_i-\sqrt{\xi_W}m_W] \nonumber \\
  &-&\frac{3 e^4 m_W}{16384\pi^2 c_w^4 s_w^4 x_t^{3/2}\xi_W^2}D^3\,
  \theta[\sqrt{\xi_W}-1/c_w-1]\sum_{i=d,s,b}\frac{|V_{3i}|^2}{B_i}\bigl{(}
  x_t^4-(3\xi_W+4 x_i)x_t^3 \nonumber \\
  &+&3(\xi_W^2+x_i\xi_W+2 x_i^2)x_t^2-(\xi_W^3-2 x_i\xi_W^2-3 x_i^2\xi_W+4 x_i^3)x_t
  -(\xi_W-x_i)^3 x_i \bigr{)}\theta[m_t-m_i-\sqrt{\xi_W}m_W]\,.
\eeqa
For the contribution of the 5th cut of Fig.5 to Eq.(5) we obtain the same result as Eq.(23). One can also see this point from the left-and-right symmetry of the 3th and 5th cuts of Fig.5.

Summing up all of the above results we obtain the $\xi_W$-dependent contribution of the cuts of Fig.3 to Eq.(5). The result has been shown in Eq.(13). From Eqs.(19-23) one also readily see that the total contribution of CKM matrix elements of each topology of Fig.3 to Eq.(5) is real number.

\end{document}